\documentstyle{article}
\begin{document}

\author{Gunn Quznetsov \\
quznets@geocities.com}
\title{THE PROBABILITY IN THE RELATIVISTIC $\mu +1$ SPACE-TIME}
\maketitle

\begin{abstract}
The probability behavior in the $\mu +1$ relativistic space-time is
considered. The probability, which is defined by the relativistic $\mu +1$%
-vector of the probability density, is investigated.
\end{abstract}

\section{INTRODUCTION}

There we will consider the probability behavior in the $\mu +1$ dimensional
relativistic space-time for to find the conditions, when the probability is
defined by the relativistic $\mu +1$-vector of the probability density..

\section{SIMPLICES}

Let $\Re $ be the $\mu $-dimensional euclidean space. Let \underline{$S$} be
the set of the couples \underline{$a$}$=\left\langle a_0,\overrightarrow{a}%
\right\rangle $, for which: $a_0$ is the real number and $\overrightarrow{a}%
\in $ $\Re $. That is if $R$ is the set of the real numbers then \underline{$%
S$} $=R$ $\times $ $\Re $.

If \underline{$a$} $\in $ \underline{$S$}, \underline{$a$}$=\left\langle a_0,%
\overrightarrow{a}\right\rangle $, $R^\mu $ is the coordinates system on $%
\Re $ and $a_1,...,a_\mu $ are the coordinates of $\overrightarrow{a}$ in $%
R^\mu $ then let us denote:

\[
\underline{a}\left( R^{\mu +1}\right) =\left\langle a_0,\overrightarrow{a}%
\right\rangle \left( R^{\mu +1}\right) =\left\langle a_0,(a_1,...,a_\mu
)\right\rangle , 
\]

''\underline{$x_1$}$+\underline{x_2}=$\underline{$x_3$}'' : ''for all $i$:
if $0\leq i\leq \mu $ then

$x_{1,i}+x_{2,i}=x_{3,i}$'',

and for every real number $k$:

''$k\cdot \underline{x_1}=$\underline{$x_2$} '' : ''for all$i:$ if $0\leq
i\leq \mu $ then $x_{2,i}=k\cdot x_{1,i}$ '',

''$\smallint d\overrightarrow{x}$'': ''$\smallint dx_1...\smallint dx_\mu $%
''.

The set $M$ of the points of \underline{$S$} is the $n$-simplex with the
vertices \underline{$a_0$}$,\underline{a_1},...,$\underline{$a_n$} (denote: $%
M=\left\lceil \underline{a_0},\underline{a_1},...,\underline{a_n}%
\right\rceil $) if for all \underline{$a$}: if \underline{$a$} $\in M$ then
the real numbers $k_1,k_2,...,k_n$ exist, for which: for all $i$: if $0\leq
i\leq n$ then $0\leq k_i\leq 1,$and

\[
\underline{a}=\underline{a_0}+\sum_{r=1}^n\left( \prod_{i=1}^rk_i\right)
\cdot \left( \underline{a_{n-r+1}}-\underline{a_{n-r}}\right) . 
\]

Let $\Delta (\underline{a_0},\underline{a_1},...,\underline{a_\mu },$%
\underline{$a_{\mu +1}$}$)\left( R^{\mu +1}\right) $ be the determinant:

\[
\left| 
\begin{array}{cccc}
a_{\mu +1,0}-a_{0,0} & a_{\mu +1,1}-a_{0,1} & \cdots & a_{\mu +1,\mu
}-a_{0,\mu } \\ 
a_{\mu ,0}-a_{0,0} & a_{\mu ,1}-a_{0,1} & \cdots & a_{\mu ,\mu }-a_{0,\mu }
\\ 
\cdots & \cdots & \cdots & \cdots \\ 
a_{1,0}-a_{0,0} & a_{1,1}-a_{0,1} & \cdots & a_{1,\mu }-a_{0,\mu }
\end{array}
\right| 
\]
.

Let $v$ be some real number, for which: $|v|\leq 1.$

Let $R^{\mu +1\prime }$ is obtained from $R^{\mu +1}$ by the Lorenz
transformations:

For some $k$, for which $1<k<\mu $:

\[
a_k^{\prime }=\frac{a_k-v\cdot a_0}{\sqrt{1-v^2}},a_0^{\prime }=\frac{%
a_0-v\cdot a_k}{\sqrt{1-v^2}} 
\]

and for all other $r$ for which $1\leq r\leq \mu $ and $r\neq k$: $%
a_r^{\prime }=a_r$ .

In this case:

\[
\Delta (\underline{a_0},\underline{a_1},...,\underline{a_\mu },\underline{%
a_{\mu +1}})\left( R^{\mu +1\prime }\right) =\Delta (\underline{a_0},%
\underline{a_1},...,\underline{a_\mu },\underline{a_{\mu +1}})\left( R^{\mu
+1}\right) . 
\]

Let the $\mu +1$-measure of the $\mu +1$-simplex $\left\lceil \underline{a_0}%
,\underline{a_1},...,\underline{a_\mu },\underline{a_{\mu +1}}\right\rceil $
be:

\[
\left\| \underline{a_0},\underline{a_1},...,\underline{a_\mu },\underline{%
a_{\mu +1}}\right\| =\frac 1{\left( \mu +1\right) !}\cdot \Delta (\underline{%
a_0},\underline{a_1},...,\underline{a_\mu },\underline{a_{\mu +1}}). 
\]

The $\mu +1$-measure of $\mu +1$-simplex is invariant for the complete
Poincare group transformations.

If $M_{1,k}$ is the subdeterminant of

\[
\Delta (\underline{a_0},\underline{a_1},...,\underline{a_\mu },\underline{%
a_{\mu +1}}),
\]

obtained from 

\[
\Delta (\underline{a_0},\underline{a_1},...,\underline{a_\mu },\underline{%
a_{\mu +1}})
\]

by the crossing out of the first line and the column of number $k$ then the $%
\mu $-measure of the $\mu $-simplex 

\[
\left\lceil \underline{a_0},\underline{a_1},...,\underline{a_\mu }%
\right\rceil 
\]

in the coordinates system $R$ is:

\[
\left\| \underline{a_0},\underline{a_1},...,\underline{a_\mu }\right\|
(R^{\mu +1})=\frac 1{\mu !}\cdot \left( \sum_{k=1}^{\mu +1}\left(
M_{1,k}\right) ^2\right) ^{0.5}\mbox{.} 
\]

If $R^{\mu +1\prime }$ is obtained by the Lorentz transformations from $%
R^{\mu +1}$ and for all $k$ and $s$, for which $0\leq k\leq \mu $, $0\leq
s\leq \mu $:

$a_{k,0}=a_{s,0}$,

then

\[
\left\| \underline{a_0},\underline{a_1},...,\underline{a_\mu }\right\|
(R^{\mu +1\prime })=\left\| \underline{a_0},\underline{a_1},...,\underline{%
a_\mu }\right\| (R^{\mu +1})\cdot \sqrt{1-v^2}. 
\]

Let $S^{\#}$ be the related to \underline{$S$} vector space. That is the
couple $(\underline{S},S^{\#})$ is the affine space.

In the coordinates system $R^{\mu +1}:$

If \underline{$n$}$\in S^{\#}$ then the modulus of \underline{$n$} is:

\[
|\underline{n}|=\left| \left\langle n_0,\overrightarrow{n}\right\rangle
\right| =(n_0^2+n_1^2+...+n_\mu ^2). 
\]

The scalar product of the vectors \underline{$n_1$} and \underline{$n_2$} is
denoted as:

\[
\underline{n_1}\cdot \underline{n_2}=\left\langle n_{1,0},\overrightarrow{n_1%
}\right\rangle \cdot \left\langle n_{1,0},\overrightarrow{n_2}\right\rangle
=n_{1,0}\cdot n_{2,0}+n_{1,1}\cdot n_{2,1}+...+n_{1,\mu }\cdot n_{2,\mu }. 
\]

The vector \underline{$n$} for which: \underline{$n$}$\in S^{\#}$ and $%
n_{k-1}=(-1)^{1=k}\cdot M_{1,k}$ .is the normal vector for the $\mu $%
-simplex $\left\lceil \underline{a_0},\underline{a_1},...,\underline{a_\mu }%
\right\rceil $. (denote: \underline{$n$} $\bot $ $\left\lceil \underline{a_0}%
,\underline{a_1},...,\underline{a_\mu }\right\rceil $).

Let us denote:

\[
\partial _k=\frac \partial {\theta x_k},\partial _t=\frac \partial {\theta
t}=\partial _0\mbox{.} 
\]

Let $\Re ^{\#}$ be the related to $\Re $ the vector space. That is $(\Re
^{\#},\Re )$ is the affine space.

Let us denote: for $\overrightarrow{n}\in \Re ^{\#}$ : $\overrightarrow{n}%
^2=n_1^2+n_2^2+...+n_\mu ^2$.

Let us denote the zero vector of $\Re ^{\#}$ as the following: for $%
\overrightarrow{n}\in \Re ^{\#}$: if $\overrightarrow{n}=\overrightarrow{0}$
then for all $k$: if $1\leq k\leq \mu $ then $n_k=0$.

Let us denote the vector \underline{$e$}$(\underline{S})$ as the basic
vector of $\underline{S}$ if \underline{$e$}$(\underline{S})=\left\langle 1,%
\overrightarrow{0}\right\rangle \in S^{\#}$.

TRACKS

Let a differentiable real vector function $\overrightarrow{f}(t)$ ($%
\overrightarrow{f}(t)\in \Re ^{\#}$) be denoted as the track in $R^{\mu +1}$
.

Let the distance between the tracks $\overrightarrow{f_1}$ and $%
\overrightarrow{f_2}$ be denoted as the following:

\[
\left\| \overrightarrow{f_1},\overrightarrow{f_2}\right\| =\sup_t\left(
\left( \sum_{i=1}^\mu \left( f_{1,i}(t)-f_{2,i}(t)\right) ^2\right)
^{0.5}\right) \mbox{.} 
\]

$\left\| \overrightarrow{f_1},\overrightarrow{f_2}\right\| $ fulfilles to
all three metric space axioms:

1) $\left\| \overrightarrow{f_1},\overrightarrow{f_2}\right\| $=0 and if $%
\overrightarrow{f_1}\neq \overrightarrow{f_2}$ then $\left\| \overrightarrow{%
f_1},\overrightarrow{f_2}\right\| >0$;

2) $\left\| \overrightarrow{f_1},\overrightarrow{f_2}\right\| =\left\| 
\overrightarrow{f_2},\overrightarrow{f_1}\right\| $;

3) By the Cauchy-Schwarz inequality:

$\left\| \overrightarrow{f_1},\overrightarrow{f_2}\right\| +\left\| 
\overrightarrow{f_2},\overrightarrow{f_3}\right\| \geq \left\| 
\overrightarrow{f_1},\overrightarrow{f_3}\right\| $.

In this case the set $T$ of the tracks in $R^{\mu +1}$ is the metric space.
The topology on the set $T$ can be constructed by the following way:

Let the set $O_\varepsilon (\overrightarrow{f_0})$ ($O_\varepsilon (%
\overrightarrow{f_0})\subset T$) be the $\varepsilon $-vicinity of $%
\overrightarrow{f_0}$ if for all $\overrightarrow{f}$: if $\overrightarrow{f}%
\in O_\varepsilon (\overrightarrow{f_0})$ then $\left\| \overrightarrow{f},%
\overrightarrow{f_0}\right\| <\varepsilon $.

The track $\overrightarrow{f}$ is the interior point of set $M$ ($M\subseteq
T$) if $\overrightarrow{f}\in M$ and for some $\varepsilon $-vicinity $%
O_\varepsilon (\overrightarrow{f})$ of $\overrightarrow{f}$: $O_\varepsilon (%
\overrightarrow{f})\subseteq M$.

The set $M$ is the open set if all elements of $M$ are the interi- or points
of $M$.

In this case $\widehat{B}$ can be the minimum $\sigma $-field (The Borel
field) contained all open subsets of $T$.

Let $Ptr$ be the probability measure on $\widehat{B}$. That is $(T,\widehat{B%
},Ptr)$ is the probability space.

The vector-function $[w(t_0,\left\lceil \underline{a_0},\underline{a_1},...,%
\underline{a_\mu }\right\rceil )],$ which has got the range of values in $%
\Re ^{\#},$ is

the average velocity of the tracks density on the $\mu $-simplex $%
\left\lceil \underline{a_0},\underline{a_1},...,\underline{a_\mu }%
\right\rceil $ in the moment $t_0$ if

\[
\lbrack w(t_0,\left\lceil \underline{a_0},\underline{a_1},...,\underline{%
a_\mu }\right\rceil )]= 
\]

\[
\frac{\int dy\cdot y\cdot Ptr\left( \left\{ \overrightarrow{f}:\partial _t%
\overrightarrow{f}\left( t_0\right) =y\right\} \cap \left\{ \overrightarrow{f%
}:\overrightarrow{f}\left( t_0\right) \in \left\lceil \underline{a_0},%
\underline{a_1},...,\underline{a_\mu }\right\rceil \right\} \right) }{%
Ptr\left( \left\{ \overrightarrow{f}:\overrightarrow{f}\left( t_0\right) \in
\left\lceil \underline{a_0},\underline{a_1},...,\underline{a_\mu }%
\right\rceil \right\} \right) }\mbox{.} 
\]

The vector-function $w(t,\overrightarrow{x}),$ which has got the domain in 
\underline{$S$} and has got the range of values in $\Re ^{\#}$, is the
velocity of the tracks density, if for all $k$, for which $0\leq k\leq \mu $%
: if \underline{$a_k$}$\rightarrow \left\langle t_0,\overrightarrow{x_0}%
\right\rangle $ then

\[
w(t_0,\overrightarrow{x_0})=lim[w(t_0,\left\lceil \underline{a_0},\underline{%
a_1},...,\underline{a_\mu }\right\rceil )]. 
\]

Let the real function $ptr(\underline{n},t_0,\overrightarrow{x})$ be the
tracks probability density for the direction of $\underline{n}$ for $R^{\mu
+1}$ if for all $\underline{a_0},\underline{a_1},...,\underline{a_\mu },$
for which: \underline{$n$} $\bot $ $\left\lceil \underline{a_0},\underline{%
a_1},...,\underline{a_\mu }\right\rceil $, the following condition is
fulfilled:

if for all $i$ ($0\leq i<\mu $): \underline{$a_i$}$\rightarrow \left\langle
t_0,\overrightarrow{x_0}\right\rangle $ then

\[
ptr(\underline{n},t_0,\overrightarrow{x_0})=\lim \frac{Ptr\left( \left\{ 
\overrightarrow{f}:\overrightarrow{f}\left( t\right) \in \left\lceil 
\underline{a_0},\underline{a_1},...,\underline{a_\mu }\right\rceil \right\}
\right) }{\left\| \underline{a_0},\underline{a_1},...,\underline{a_\mu }%
\right\| }\mbox{.} 
\]

Let $(\underline{n_1}\symbol{94}\underline{n_2})$ be the angle between 
\underline{$n_1$} and \underline{$n_2$} . That is: $\cos (\underline{n_1}%
\symbol{94}\underline{n_2})=(\underline{n_1}\cdot \underline{n_2})/(\left| 
\underline{n_1}\right| \cdot \left| \underline{n_2}\right| ).$ In this case,
if

\underline{$w$}$(t,\overrightarrow{x})=\left\langle w_0(t,\overrightarrow{x}%
).\overrightarrow{w}(t,\overrightarrow{x})\right\rangle \in S^{\#}$ , $w_0(t,%
\overrightarrow{x})=1$, $\overrightarrow{w}(t,\overrightarrow{x})$ is the
velocity of the tracks density and $-\frac \pi 2\leq ($\underline{$w$}$(t,%
\overrightarrow{x})\symbol{94}$\underline{$n$})$\leq \frac \pi 2$ then

\begin{equation}
ptr(\underline{n},t,\overrightarrow{x})=ptr(\underline{w}(t,\overrightarrow{x%
}),t,\overrightarrow{x})\cdot \cos \left( \underline{w}(t,\overrightarrow{x})%
\symbol{94}\underline{n}\right) .  \label{p1}
\end{equation}

\section{TRACKELIKE PROBABILITY}

Let the $\sigma $-field $\widetilde{B}$ on \underline{$S$} be obtained from
the set of the $\mu +1$-simplices. Let the probability measure $P$ on $%
\widetilde{B}$ be defined as the following:

the real function $p(t,\overrightarrow{x})$ (the absolute probability
density) exists for which:

if $D\in \widetilde{B}$ then

\[
0\leq \int dt\int \int ...\int_{\left( D\right) }dx_1dx_{2\cdots }dx_\mu
\cdot p\left( t,x_1x_2,\cdots ,x_\mu \right) \leq 1 
\]

and

\[
\int dt\int \int ...\int_{\left( R^{\mu +1}\right) }dx_1dx_{2\cdots }dx_\mu
\cdot p\left( t,x_1x_2,\cdots ,x_\mu \right) =1; 
\]

in this case:

\[
P(D)=\int dt\int \int ...\int_{\left( D\right) }dx_1dx_{2\cdots }dx_\mu
\cdot p\left( t,x_1x_2,\cdots ,x_\mu \right) . 
\]

Because for the Lorentz transformations

\[
x_k^{\prime }=\frac{x_k-v\cdot t}{\sqrt{1-v^2}}\mbox{, }t^{\prime }=\frac{%
t-v\cdot x_k}{\sqrt{1-v^2}}\mbox{,} 
\]

for $r\neq k$: $x_r^{\prime }=x_r$ ($\left| v\right| <1$),

the Jacobian:

\[
J=\frac{\partial \left( t^{\prime },x_k^{\prime }\right) }{\partial \left(
t,x_k\right) }=1 
\]

then

\[
p^{\prime }(t^{\prime },\overrightarrow{x^{\prime }})=p(t,\overrightarrow{x}%
) 
\]

That is the absolute probability density is the scalar function.

Let $g(\overrightarrow{n},t,\overrightarrow{x})$ be the conditional
probability density for the direction of $\overrightarrow{n}$, if $%
\overrightarrow{n}\in \Re ^{\#}$, \underline{$n$}$=\left\langle n_0,%
\overrightarrow{n}\right\rangle \in $\underline{$S$}$^{\#}$, $n_0=1$ and for
all points \underline{$x_0$}, for which \underline{$x_0$}$=\left\langle t_0,%
\overrightarrow{x_0}\right\rangle \in $\underline{$S$}:

\[
g(\overrightarrow{n},t,\overrightarrow{x})=\frac{p\left( t_0,\overrightarrow{%
x_0}\right) \cdot \cos \left( \underline{e}\left( \underline{S}\right) 
\symbol{94}\underline{n}\right) }{\int d\overrightarrow{x}\cdot p\left( t_0+%
\overrightarrow{n}\cdot \left( \overrightarrow{x}-\overrightarrow{x_0}%
\right) ,\overrightarrow{x}\right) }\mbox{.} 
\]

The probability measure $P$ is the trackelike probability measure in the
point \underline{$a$} (\underline{$a$}$=\left\langle t,\overrightarrow{x}%
\right\rangle \in $\underline{$S$}) if the vector $\overrightarrow{u}(t,%
\overrightarrow{x})$ exists, for which $\overrightarrow{u}(t,\overrightarrow{%
x})\in \Re ^{\#}$, and the following condition is fulfilled:

for all vectors \underline{$n$} (\underline{$n$}$=\left\langle n_0,%
\overrightarrow{n}\right\rangle \in $\underline{$S$}$^{\#}$ , $n_0=1$):

if \underline{$u$}$(t,\overrightarrow{x})=\left\langle u_0(t,\overrightarrow{%
x}),\overrightarrow{u}(t,\overrightarrow{x})\right\rangle \in $\underline{$S$%
}$^{\#}$, $u_0(t,\overrightarrow{x})=1$

and $-\frac \pi 2\leq \left( \underline{u}(t,\overrightarrow{x})\symbol{94}%
\underline{n}\right) \leq \frac \pi 2$

then (see (\ref{p1})):

\[
g(\overrightarrow{n},t,\overrightarrow{x})=g(\overrightarrow{u}(t,%
\overrightarrow{x}),t,\overrightarrow{x})\cdot \cos \left( \underline{u}(t,%
\overrightarrow{x})\symbol{94}\underline{n}\right) . 
\]

In this case $\overrightarrow{u}(t,\overrightarrow{x})$ is denoted as the
velocity of the probability in the point $\left\langle t,\overrightarrow{x}%
\right\rangle $.

If $P$ is the trackelike probability measure in the point \underline{$a_0$} (%
\underline{$a_0$}$\in $\underline{$S$} and \underline{$a_0$}$=\left\langle
t_0,\overrightarrow{x_0}\right\rangle $) and $u_0=1=n_0$ then

\[
\frac{\int d\overrightarrow{x}\cdot p\left( t_0+\overrightarrow{u}(t_0,%
\overrightarrow{x_0})\cdot \left( \overrightarrow{x}-\overrightarrow{x_0}%
\right) ,\overrightarrow{x}\right) }{\left( \cos \left( \underline{u}\left(
t_0,\overrightarrow{x_0}\right) \symbol{94}\underline{n}\right) \cdot \cos
\left( \underline{u}\left( t_0,\overrightarrow{x_0}\right) \symbol{94}%
\underline{e}\left( \underline{S}\right) \right) \right) }= 
\]

\begin{equation}
=\frac{\int d\overrightarrow{x}\cdot p\left( t_0+\overrightarrow{u}(t_0,%
\overrightarrow{x_0})\cdot \left( \overrightarrow{x}-\overrightarrow{x_0}%
\right) ,\overrightarrow{x}\right) }{\cos \left( \underline{u}\left( t_0,%
\overrightarrow{x_0}\right) \symbol{94}\underline{e}\left( \underline{S}%
\right) \right) }\mbox{.}  \label{p2}
\end{equation}

If $q(t,\overrightarrow{x})=g(\overrightarrow{0},t,\overrightarrow{x})$ then 
$\rho (t,\overrightarrow{x})$ is the density function in the moment $t$.

If $\overrightarrow{u}(t,\overrightarrow{x})$ is the velocity of the
probability, then the function $\overrightarrow{j}(t,\overrightarrow{x})$,
which has got the domain in $\underline{S}$ and has got the range of values
in $\Re ^{\#}$, is denoted as the probability current if

\[
\overrightarrow{j}(t,\overrightarrow{x})=\rho (t,\overrightarrow{x})\cdot 
\overrightarrow{u}(t,\overrightarrow{x}). 
\]

These function are fulfilled to the continuity equation:

\[
\partial _t\rho (t,\overrightarrow{x})+\partial _1j_1(t,\overrightarrow{x}%
)+\cdots +\partial _\mu j_\mu (t,\overrightarrow{x})=0. 
\]

Let $u$ be the velocity of the probability in the point $\left\langle t_0,%
\overrightarrow{x_0}\right\rangle $ and the coordinates system $R^{\mu
+1\prime }$ be obtained from the coordi- nates system $R^{\mu +1}$ by the
Lorentz transformations with the velo- city $u$. That is:

\[
t^{\prime }=\frac{t-\overrightarrow{u}\cdot \overrightarrow{x}}{\sqrt{1-%
\overrightarrow{u}^2}}\mbox{ and }\overrightarrow{x}^{\prime }=\frac{%
\overrightarrow{x}-t\cdot \overrightarrow{u}}{\sqrt{1-\overrightarrow{u}^2}}%
\mbox{.} 
\]

In this case $\rho ^{\prime }(t^{\prime },\overrightarrow{x}^{\prime })$ is
denoted as the local probability

density ($\rho _{\bigcirc }$ ). This function is the scalar function:

\[
\rho ^{\prime }(t^{\prime },\overrightarrow{x}^{\prime })=\rho (t,%
\overrightarrow{x}); 
\]

and

\[
\rho (t,\overrightarrow{x})=\frac{\rho _{\bigcirc }(t,\overrightarrow{x})}{%
\sqrt{1-\overrightarrow{u}^2\left( t,\overrightarrow{x}\right) }}\mbox{.} 
\]

Hence for any velocity $v$, for which $|v|<1$: if

\[
t^{\prime }=\frac{t-\overrightarrow{v}\cdot \overrightarrow{x}}{\sqrt{1-%
\overrightarrow{v}^2}}\mbox{ and }\overrightarrow{x}^{\prime }=\frac{%
\overrightarrow{x}-t\cdot \overrightarrow{v}}{\sqrt{1-\overrightarrow{v}^2}} 
\]

then

\[
\rho ^{\prime }(t^{\prime },\overrightarrow{x}^{\prime })=\frac{\rho (t,%
\overrightarrow{x})-\overrightarrow{v}\cdot \overrightarrow{j}(t,%
\overrightarrow{x})}{\sqrt{1-\overrightarrow{v}^2}}\mbox{,} 
\]

\[
\overrightarrow{j}^{\prime }(t^{\prime },\overrightarrow{x}^{\prime })=\frac{%
\overrightarrow{j}(t,\overrightarrow{x})-\rho (t,\overrightarrow{x})\cdot 
\overrightarrow{v}}{\sqrt{1-\overrightarrow{v}^2}}\mbox{.} 
\]

Therefore $\rho (t,\overrightarrow{x})$ is not the scalar function but:

\[
\rho ^2(t,\overrightarrow{x})-\overrightarrow{j}^2(t,\overrightarrow{x}%
)=\rho _{\bigcirc }^2(t,\overrightarrow{x}). 
\]

\section{RESUME}

In order to the probability is defined by the relativistic $\mu +1$-vector
of the density, the probability distribution function must fulfil to the odd
global condition (\ref{p2}), which is expressed by the integrals on all
space.

\end{document}